\documentclass[amsmath,amssymb,floatfix,reprint,twocolumn,superscriptaddress,prb]{revtex4-2}

\usepackage{graphicx}
\usepackage{epsfig}
\usepackage{hyperref}
\usepackage{xcolor}
\usepackage{bm}

\begin{document}

\title{Sample shape dependence of magnetic noise}
\author{Steven T. Bramwell }
\date{\today}

\affiliation{
London Centre for Nanotechnology and Department of Physics \& Astronomy, University College London,
17-19 Gordon Street, London, WC1H 0AJ, United Kingdom}

\begin{abstract}
Zero-field magnetic noise, characterised by the magnetic autocorrelation function $S_s(t)$, has been observed, perhaps surprisingly, to depend on sample shape $s$. The reasons for this are identified and general expressions are derived that relate the autocorrelation functions for systems of different shape to an underlying `intrinsic' form. Assuming the flcutuatiopn-dissipation theorem, it is shown that, for any noise that relaxes monotonically, the effect of sample shape is to reduce both the noise amplitude and mean relaxation time by a factor of $1+N\chi_i$, where $N$ is the demagnetizing factor and $\chi_i$ the intrinsic susceptibility. In frequency space, where $S_s(t)$ Fourier transforms into the power spectrum $S_s(\omega)$, the above two factors combine to suppress the zero frequency amplitude of $S_s(\omega)$ by $(1+N\chi_i)^2$, while at high frequency, sample shape dependence becomes negligible. These results suggest simple and robust experimental tests of the fluctuation--dissipation theorem in magnetic systems that may be useful in distinguishing bulk from surface effects. 
\end{abstract}

\maketitle

Magnetic noise measurements add an extra dimension to the study of magnetic correlations that complement  more conventional methods: they have been applied in the past to spin glasses~\cite{Ocio, Reim, spinglass}, and more recently have been discussed and applied in the context of spin ice~\cite{K,D,S,H,Morineau,B}. 
In a recent experimental study of the fluctuation-dissipation theorem, using spin ice as a model system, Morineau {\it et al.}~\cite{Morineau} measured~\cite{note} the autocorrelation function $S_s(t)$ of magnetic noise. They observed that $S_s(t)$ depends on sample shape (here denoted $s$), and commented that this seems surprising in view of the fact that noise is measured at zero applied field. 

To expand on this point, magnetic noise clearly does depend on $s$ as the power spectrum $S_s(\omega)$, the Fourier transform~\cite{Fourier} of $S_s(t)$, depends on the imaginary part of the complex susceptibility $\chi_s(\omega)$, by means of the fluctuation--dissipation theorem; and $\chi_s(\omega)$ depends on sample shape, via the demagnetizing factor $N$, in the same way as the static susceptibility~\cite{Finger}. But this does seem surprising because $N$ only enters magnetic problems as a correction to the applied field $H$, which is zero in the noise measurements~\cite{Morineau}. One might also add that $N$ does not explicitly enter the Hamiltonian of system $s$, so cannot affect its dynamics.  Also, the complicated manner in which $N$ affects $S_s(\omega)$ (see below) does not offer immediate insight into how, or why, $S_s(t)$ varies with $N$. 

It therefore seems worthwhile asking, why does $S_s(t)$ depend on $N$ and 
is there a general transformation that relates $S_s(t)$ for systems of different shape $s$?  Answering these questions starts with an articulation of the results of Ref. \cite{Finger} which introduces a shape--independent reference system $i$ that coincides with physical systems for which $N=0$. It can be shown~\cite{Finger} that the Hamiltonians of systems $s$ and $i$ differ by the energy $(\mu_0/2)V NM(t)^2$ (here $M$ is the magnetization, $\mu_0$ the vacuum permeability and $V$ the volume). The relationship may be written:
\begin{equation}\label{Ham}
\mathcal{H}_s = \mathcal{H}_i - (\mu_0/2)V NM(t)^2,  
\end{equation}
where the term in $M(t)$ acts as a classical number, so does not affect the dynamics, but does affect statistical weights. In finite field there is the relation~\cite{Finger}: 
\begin{equation}\label{one}
\rho_s[H(t),t] = \rho_i[H_{\rm in}(t),t], 
\end{equation}
where $\rho$ is a density matrix and $H_{\rm in}(t) = H(t)-N M(t)$ is the time-dependent internal field. 

One can see from Eq. \ref{Ham}, that even though $N$ does not appear in $\mathcal{H}_s$, it enters the problem if we want to compare system $s$ with system $i$. This comparison is most easily performed at finite field using Eq. \ref{one}, so, for the purpose of comparing $S_s(t)$ with $S_i(t)$, it expedient to assume the fluctuation--dissipation theorem: a spontaneous thermal fluctuation in magnetization relaxes according to the same equations of motion as does an equilibrium magnetization prepared by applying a field, and then removing it at $t=0$~\cite{Onsager,CW, Kubo}. Eq. \ref{one} then maps the properties of a system of shape $s$ onto those of the reference system $i$. In practical terms this enables the identification of intrinsic or shape-independent properties via measurements on samples of regular shapes for which $N$ is known. These include spheroids, cuboids, cylinders etc. (note that for non--spheroids, $N$ is a function of the static intrinsic susceptibility $\chi_i$~\cite{Chen, Twengstrom1, Twengstrom2}).  However on the right hand side of the equation, $H_{\rm in}(t)$ is not known a-priori, while $H(t)$ is a controlled variable, so the key to using Eq. \ref{one} is to express $H_{\rm in}(t)$ in terms of the known $H(t)$. 

It should be noted that the results of Ref~\cite{Finger} are derived by neglecting magnetostatic modes and assuming that magnetization fluctuations  can be represented in terms of normal modes ${\bf M}({\bf q})$.  This idealisation is appropriate for zero field noise in highly correlated paramagnets, the main subject of interest here.  The magnetization ${\bf M}({\bf 0}) \equiv M$ is the only mode that couples to an applied field $H$, with which it aligns (hence both are treated as scalars here) and is the only one affected by sample shape, via the demagnetizing field $-N M$~\cite{Finger}.\\

\noindent
{\it Why $S_s(t)$ is shape dependent: --}

The normal mode description ensures that the system of shape $s$ does have precisely the same, shape-independent, dynamics as that of the reference system $i$. But what differs, is any statistically averaged property that involves the magnetization $M$, which includes the autocorrelation function $S(t) = \mu_0 V \langle M(0)M(t) \rangle$. As a heuristic approach \cite{Lenk} we consider that if the system is in a microstate with a particular value of $M$ at some particular time, the average value of $M$ reached at some later time does not depend on $s$. It follows that 
\begin{equation}\label{x}
S_s(t) = \mu_0 V \langle M e^{-M H(t) /kT} f^M(t)\rangle_s. 
\end{equation}
where the function $f$ does not depend on $s$, but the canonical average $\langle\dots\rangle_s$ does. 
Applying Eq. \ref{one}, and contrasting with the reference system, we find: 
\begin{equation}\label{two}
S_s(t) = \mu_0 V \langle M e^{-M H_{\rm in}(t) /kT} f^M(t)\rangle_i, 
\end{equation}
\begin{equation}\label{three}
S_i(t) = \mu_0 V \langle M e^{-M H(t) /kT} f^M(t)\rangle_i. 
\end{equation}
Comparison of Eqs. \ref{x}, \ref{two}, \ref{three} shows why $S_s(t)$ depends on sample shape. In zero field, fluctuations in magnetization cost a different amount of energy that makes $\langle\dots\rangle_s$ different to $\langle\dots\rangle_i$. Such fluctuations relax as if a field had been removed, but if one wants to use the canonical distribution of $i$ to calculate this, then the field removed must be set to a fictitious value that depends on $N$ (Eqs. \ref{two}, \ref{three}).  The same dynamics (function $f^M(t)$) then leads to a different $S_s(t)$.  This seems to answer the question posed in Ref. \cite{Morineau} of how the zero field property $S_s(t)$ can be shape--dependent. \\

\noindent
{\it Linear response calculation: - }

The actual time dependencies can be calculated in linear response theory, where a `feedback' between between field, magnetization and autocorrelation function becomes evident, even when $H=0$. To review some basic classical relationships (for the quantum theory see e.g. Ref. \cite{ML}), a statement of the fluctuation--dissipation theorem is that $S(t)$ (minus any time--independent part) is related to the  relaxation function $R(t)$ by: 
\begin{equation}
S(t)=kT R(t)
\end{equation}
where $R(t)$ is considered to be an even function of time~\cite{even}. With this definition, $R(0) = \chi$, the static susceptibility. The relaxing magnetization for the field protocol of interest is, for positive times, 
$
M(t) = HR(t).
$
For example, given $H(t) = H$ for $t<0$ and $H(t) = 0$, $R(t) = \chi e^{-\lambda t}$ for $t>0$, then $M(t)$ would take the form $M(t) =\chi H$ and  $\chi H e^{-\lambda t}$ respectively. The time--dependent susceptibility is defined to be zero for negative $t$ and is equal to $-\partial R(t)/\partial t$ for positive $t$. In the previous example, it would take the form $\chi(t) = 0$ and $\chi \lambda e^{-\lambda t}$ respectively (note that $\chi(0) \ne \chi$). Now, because $M(t)$ at time $t$ is influenced by $H(t')$ at earlier times ($t'< t$), the general relation between a change in magnetization $\delta M(t)$ and a change in field $\delta H(t)$ is a convolution rather than a simple product: 
$ \delta M(t) = \chi(t) \ast \delta H(t)$. 

Returning to the two systems $s$ and $i$, a comparison of the time dependencies of $S_i(t)$ and $S_s(t)$ amounts to comparing $R_i(t)$ and $R_s(t)$ respectively, and this requires a relation between $\chi_i(t)$ and $\chi_s(t)$. Such a relation can be derived using Eq. \ref{one}~\cite{Finger}: 
\begin{equation}\label{solve}
\chi_s(t) = \chi_i(t) - N \chi_s(t)\ast \chi_i(t). 
\end{equation}
By solving this equation one can relate the $R$'s and hence relate $S_i(t)$ to $S_s(t)$. \\

\noindent
{\it Transformations of the autocorrelation functions: - }

A straightforward way of solving the above equations is to Laplace (or Fourier) transform Eq. \ref{solve} to give:
$
\chi_s(\omega) = \chi_i(\omega) - N \chi_s(\omega) \chi_i(\omega),
$
and then to isolate the imaginary part, giving:
\begin{equation}\label{KK}
\chi''_{i/s}(\omega) =
\frac{\chi''_{s/i}(\omega)}
{N^2 \chi''_{s/i}(\omega)^2 +\left(N \mathcal{K}[\chi''_{s/i}(\omega)]\mp1 \right)^2}
\end{equation}
where $-$ and $+$ on the right hand side correspond to $i$ and $s$ respectively, and $\mathcal{K}[\chi''(\omega)] = \chi'(\omega)$ is the Kramers-Kronig inversion of $\chi''(\omega)$. The transformation for the $S_{i/s}(\omega)$ can then be found using $S_{i/s}(\omega) = 2 k T |\chi''_{i/s}(\omega)|/\omega$, a familiar form of the fluctuation dissipation theorem which can be derived from the linear response equations given above. This gives, in principle,  $S_{i/s}(t)$ as the inverse Fourier transform of $S_{i/s}(\omega)$.

Eq. \ref{KK} is a complicated transformation that can be tricky to apply in practice. As a more transparent alternative to solving Eqn. \ref{solve}, denote $\chi_{i/s}(t)$ as $g/f$ respectively and develop the iterative series:
$
f=g -N g*g + N^2 g*g*g +\dots 
$
which can be written: 
\begin{equation}\label{s1}
f= \sum_{n=0}^{\infty}(-1)^{n-1} N^{n-1} g^{(n)}
\end{equation}
where $g^{n}$ is the $n$'th `self convolution' of $g$. There is a similar series for $g$:
\begin{equation}\label{s2}
g=\sum_{n=0}^{\infty} N^{n-1} f^{(n)}.
\end{equation}
These two equations give opportunities to explore transformations of the autocorrelation function without taking a Fourier transform. \\

\noindent
{\it The case of exponential relaxation: --}

If the intrinsic relaxation function is $R_i(t; \lambda_i)=\chi_i e^{-\lambda_i t}$ then 
$g=\lambda_i\chi_i  e^{-\lambda_i t}$. The self-convolution integrals in Eq. \ref{s1} (say) are easily performed term by term and the infinite series re-summed to give: 
$
f= \lambda_i \chi_i e^{-\lambda_i(1+N\chi_i) t}. 
$
Integrating this expression gives 
$
R_s(t) = \left(\chi_i/(1+N\chi_i)\right) e^{-\lambda_i (1+N\chi_i)t } = \chi_s e^{-\lambda_i (1+N\chi_i)t }. 
$
Including a similar analysis for Eq. \ref{s2}, the final transformations (that also apply to the $R$'s) are: 
\begin{equation}\label{t1}
R_{i}(t; \lambda) = \alpha_s^{-1} R_{s}(t; \alpha_s\lambda_s)
\end{equation}
\begin{equation}\label{t2}
R_{s}(t; \lambda_s) =  \alpha_i^{-1} R_{i}(t; \alpha_i\lambda_i)
\end{equation}
where $\lambda_s = \alpha_i \lambda_i $, $\lambda_i = \alpha_s \lambda_s $, $\alpha_i = 1+N\chi_i$,  $\alpha_s = 1- N\chi_s$ and $\alpha_i\alpha_s=1$. Corresponding transformations are (respectively) 
$
\chi_{i/s}(t; \lambda_{i/s}) =  \chi_{s/i}(\alpha_{s/i} \lambda_{s/i} t))
$
and
$
S_{i/s}(\omega;\lambda_{i/s}) =  \alpha_{s/i}^{-2} S_{s/i}(\omega; \alpha_{s/i}\lambda_{s/i}),
$
where the exponent $-2$ in the latter is easily established by considering the Fourier transform of $S_s(t)$. 

The above results establish that, for exponential noise, the effect of sample shape is to reduce the relaxation time $1/\lambda$  by a factor $1+N \chi_i$, while also suppressing the zero frequency amplitude of the noise by the same factor. In frequency space, the power spectrum is reduced in amplitude, by $(1+ N\chi_i)^2$, but `stretched' on the frequency axis by $1+N\chi$. These transformations are illustrated in Fig. 1. 
\begin{figure}[!ht]
	\centering
	\includegraphics[width=0.9\linewidth]{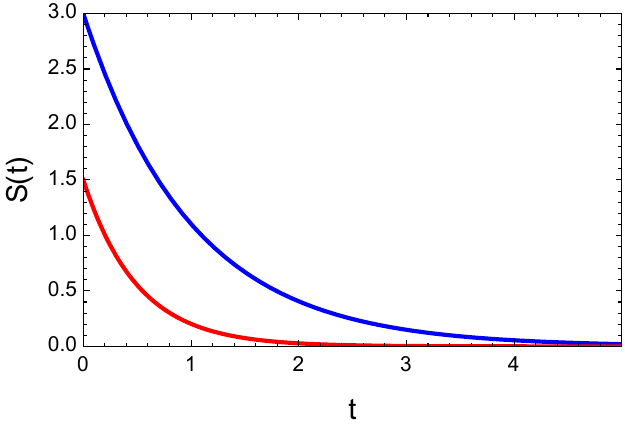}
 	\includegraphics[width=0.9\linewidth]{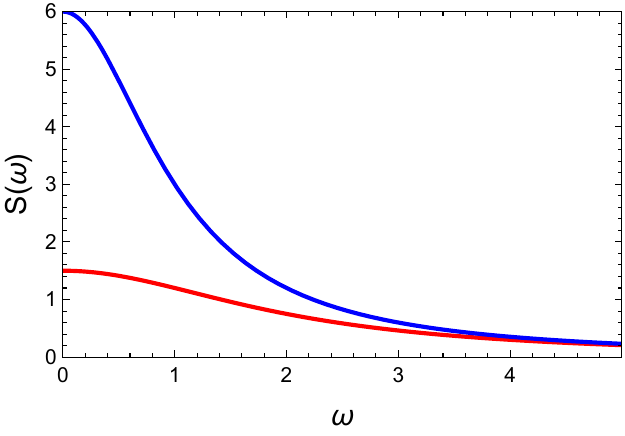}
	\caption{Shape dependence of magnetic noise for intrinsically exponential relaxation $R_i(t) = \chi_i e^{-t}$. 
    Here $kT$ is set to unity, $\chi_i=3$, and for system $s$, $N=1/3$. 
    Upper plot: intrinsic autocorrelation function $S_i(t)$ (blue, upper curve, with $S_i(t=0) = \chi_i$) and shape-dependent autocorrelation function $S_s(t)$ (red, upper curve, with  $S_s(t=0) = \chi_i/(1+ N \chi_i)$). Lower plot: the corresponding power spectra $S_i(\omega)$ (blue, lower curve, with $S(\omega=0)= 2 \chi_i$) and $S_s(\omega)$ (red, lower curve, with $S_s(\omega = 0)= 2 \chi_i/(1+N \chi_i)^2$).} 
\end{figure}

Eqs. \ref{t1} and \ref{t2} do not apply exactly to any function other than exponential, but for a relaxation composed of exponential components they always represent a partial summation of the exact series \ref{s1} and \ref{s2} that neglects cross terms between the different exponentials: they are a first approximation at short times or high frequencies. \\

\noindent
{\it The case of general monotonic relaxation: - }

The result for exponential relaxation prompts us to focus on how relaxation rate or relaxation time transforms more generally for finite $N$ and it turns out that a completely general result may be derived from the iteration series Eqs.\ref{s1} and \ref{s2}. 

Consider a monotonic relaxation $R_{i/s} = \chi_{i/s} r_{i/s}(t)$  where the $r$'s are normalised to unity at $t=0$ and hence are distributions for $t\ge 0$. Then define $c_{i/s}(t) = -dr_{i/s}(t)/dt$ which behave as density functions with integrals
\begin{equation}
\tau_{i/s} = \int_0^{\infty} t \, c_{i/s}(t) \,dt,  
\end{equation}
that define the mean relaxation time. 
The time--dependent susceptibilities become 
$
\chi_{i/s}(t) = \chi_{i/s} c_{i/s}(t) 
$
and the iteration series Eqs. \ref{s2} (for example) may be then multiplied by time and integrated. The additive properties of mean values under convolution shows that the mean of the $n$th convolution $c^{(n)}$ is just $n$ times the mean of $c$, making Eq. \ref{s2} become $\chi_i \tau_i = \chi_s \tau_s \sum_{n=1}^\infty n (N \chi_s)^{n-1}$. This allows the series again to be exactly re-summed using a formula of the type $\sum_{1}^{\infty} n x^{n-1} = 1/(1-x)^2$, and after some algebra one finds the simple result:\\
\begin{equation}\label{main}
\tau_i= \alpha_s^{-1} \, \tau_s
\end{equation}
\begin{equation}\label{mainb}
\tau_s= \alpha_i^{-1} \, \tau_i
\end{equation}

These equations are the generalisation of the result for exponential relaxation derived above. The mean relaxation time is reduced by a factor of $1+ N\chi_i$.  
We may combine this result with the relation between the static susceptibilities $\chi_i$ and $\chi_s$ (i.e.$\chi_{i/s}=\chi_{s/i}/(1\mp N\chi_{s/i})$) that define the amplitude of the relaxation $R_{s/i}(t=0)$, and from this we reach the following conclusion.  {\it For any noise that relaxes monotonically, the effect of sample shape is to reduce both the noise amplitude and mean relaxation time, by a factor of $1+N\chi_i$, where $N$ is the demagnetizing factor and $\chi_i$ the intrinsic susceptibility.} 
For example, the average relaxation time and amplitude of fluctuations measured on a spherical system ($N=1/3$) of spin ice ($\chi \approx 4 K/T$) would both be roughly halved at $T=1.3$ K with respect to the intrinsic relaxation. 

This result has immediate practical consequences. If time dependent measurements are available, then the mean relaxation time can be determined by: 
\begin{equation}
\tau_s = S_s(0)^{-1} \int_0^\infty t\, d S_s(t)
\end{equation}
and Eq. \ref{main}, $\tau_s=\tau_i (1 -  N\chi_s)$, can be tested by plotting $\tau_s$ as a function of $N\chi_s$. 
Or, as a minimal approach, one could measure along only two directions ($s=1,2$) of an asymmetric crystal, eliminate $\tau_i$ from the equations, and 
test the relation
\begin{equation}\label{main2}
\chi_i = \frac{\tau_2-\tau_1}{N_1 \tau_1-N_2\tau_2},
\end{equation}
which requires only an additional measurement of the static susceptibility. \\

\noindent
{\it Analogous result in frequency space: - }

To explore the consequences of the above result in frequency space~\cite{Carley}, the Fourier transform of the susceptibility may be expanded in powers of $i \omega$: 
$
\chi_{i/s}(\omega) = \chi_{i/s} \left(1 -  i \omega \tau_{i/s} + \dots \right).  
$
Here the imaginary term in $\omega$ defines the first term in the expansion of the imaginary susceptibility, $\chi''(\omega)$, and because $S_s(\omega) = (2/\omega)|\chi''(\omega)|$ we have:
$$
S_{i/s}(\omega\rightarrow 0) = 2 \chi_{i/s} \tau_{i/s},
$$
and from this follows 
$$S_{s}(\omega\rightarrow 0) = \frac{2 \chi_{i} \tau_{i}  }{\left(1+N\chi_i \right)^2}. $$
We see that, surprisingly, the result for the mean relaxation time does not manifest in the frequency dependence of $S_s(\omega)$ but rather in its $\omega\rightarrow 0$ amplitude, where two factors of $1+ N \chi_i$, arising from the amplitude of $S_s(t)$ and mean relaxation time respectively, combine to reduce the $\omega \rightarrow 0$ amplitude of $S_s(\omega)$ by $(1+ N \chi_i)^2$. This is relevant to experiment as it is often necessary to measure noise by lock--in detection in which case the `raw' measurements are already in frequency space. The $1/(1+ N \chi_i)^2$ suppression of the amplitude of $S_s(\omega)$, already established for the exponential case, is thus a further universal result for monotonic relaxation, that combines both the universal $1+N\chi_i$ suppression of the static susceptibility and the universal $1+N\chi_i$ suppression of the mean relaxation time. In fact, the $1/(1+ N \chi_i)^2$ dependence is already evident in Eq. \ref{KK} as, in the limit $\omega \rightarrow 0$, the imaginary susceptibility tends to zero while $S_i(\omega) = 2 \chi_i''(\omega)/\omega$ remains finite. \\

As regards the frequency dependence of $S_s(\omega)$, because $\chi(\omega)$ is small at high frequencies, the effect of $N$ becomes negligible, and one can expect that $S_s(\omega) \rightarrow S_i(\omega)$ in the high frequency limit. Thus the function $S_s(\omega)$, having the smaller $\omega \rightarrow 0$ amplitude, must be `stretched' along the frequency axis in order to match $S_i(\omega)$ at high frequency. Although the stretching factor is approximately $1+\chi_i N$, as implied by the exponential approximation discussed above, this is not exact in the general case.  \\

\noindent
{\it The example of a pink noise with exponent $3/2$: - }

The case of a pink noise with exponent $3/2$ is of physical interest in the context of spin ice~\cite{H}, and to a large extent can be treated analytically, giving an illustration of several of the points established above. 
Assume the autocorrelation function: 
\begin{equation}\label{pink}
S_i(\omega) = \frac{3 \sqrt{3} k T \chi_i}{4(1+|w|^{3/2})}, 
\end{equation}
where $\omega$ is here measured in units of some characteristic unit frequency and the numerical factors are chosen to ensure that $S_i(t
\rightarrow 0) = k T \chi_i$, in which case $\tau_i = 3\sqrt{3}/8$~\cite{st}.  
Then it follows that $\chi_i'' = (3/8)\sqrt 3 \chi_i/(1+ |w|^{3/2})$, a function that can be exactly Kramers--Kronig inverted to give $\chi'_i(\omega)$.   Applying Eq.\ref{KK}, we find that 
$$
S_s(\omega) = \frac{S_i(\omega)}{N^2 \chi_i^2f_1+\left(1+\frac{N \chi_i (f_2+f_3)}{f_4}\right)^2}
$$ 
where 
$$
f_1=\frac{27 | \omega | ^2}{64 \left(| \omega | ^{3/2}+1\right)^2},
$$
$$
f_2=\pi  \left(-3 \sqrt{3} | \omega | ^{5/2}+3 \sqrt{3} | \omega | ^{11/2}+8 | \omega | ^2-8\right),
$$
$$
f_3=6 \sqrt{3} | \omega | ^4 \log (| \omega | );
$$
$$
f_4=8 \pi  \left(| \omega | ^6-1\right). 
$$
These complicated equations vividly illustrate that there is no simple transformation between $S_i(\omega)$ and $S_s(\omega)$ but, as expected, they confirm that the ratio $S_s(\omega)/S_i(\omega)$ tends to $1/(1+N \chi_i)^2$ and $1$ in the limits of small and large $\omega$ respectively. 
The actual curves are illustrated in Fig. 2 using $N=1/3$ and susceptibility $\chi_i = 4$,  representing a spherical sample of spin ice at $T  \approx 1$ K. 
\begin{figure}[!ht]
	\centering
	\includegraphics[width=0.85\linewidth]{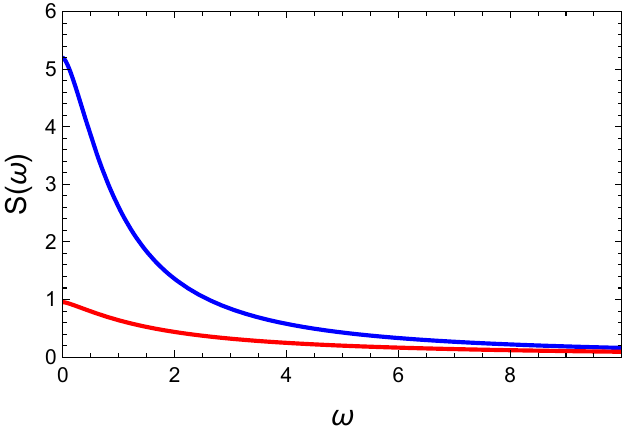}
 	\includegraphics[width=0.92\linewidth]{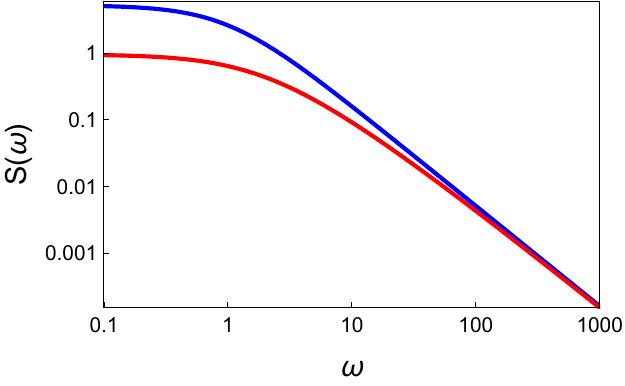}
	\caption{Sample shape dependence of the power spectrum $S(\omega)$ of magnetic noise for pink noise with exponent $3/2$ (Eq. \ref{pink}. Here $kT$ is set to unity, $\chi_i=3$, and for system $s$, $N=1/3$. 
    Upper plot: intrinsic power spectrum $S_i(t)$ (blue, upper curve) and sample shape-dependent autocorrelation function $S_s(t)$ (red, upper curve, with  $S_s(\omega=0) = S_i(\omega = 0)/(1+ N \chi_i)^2$). Lower plot: the corresponding plots on logarithmic scales, showing how $S_s(\omega) \rightarrow S_i(\omega)$ in the high frequency limit. Parameters approximately represent spin ice at $T=1$ K.} 
\end{figure}
\\

\noindent
{\it Discussion: - }

The statement derived above for $S_s(t)$ appears to be the most general one that can be made about how magnetic noise depends on sample shape. In contrast, the sample shape dependence of the functional form of the relaxation cannot be summarised in such a concise way. The effect of finite $N$ is always to suppress (in amplitude) and to compress (in time) the form of $S_s(t)$ in a universal way, but the evolution with $N$ of the functional form of the relaxation is not universal.  It appears that the exponential is the only meaningful relaxation that retains the same functional form at finite $N$. 

It is interesting to discuss these results in the context of the fluctuation--dissipation theorem, tests of which~\cite{Morineau} inspired the study. The theorem may be directly tested on a sample of shape $s$ by comparing measurements of noise with measurements of frequency--dependent susceptibility~\cite{Morineau}. A sample--shape correction is not required to confirm the applicability of the theorem to system $s$, but if a breakdown is observed, this could have its origin in either bulk or surface effects and these can only be distinguished by studying the breakdown as a function of demagnetizing factor $N$. Based on current understanding~\cite{note3} one would have to correct both the autocorrelation function and the imaginary susceptibility for finite demagnetizing factor, but this is challenging in practice. Difficulties include:  measuring $S_s(t)$ and $\chi_s(\omega)$ on an absolute scale, determining the non-adiabatic part of $Re[\chi_s(\omega)]$ (which is used in place of the Kramers-Kronig inversion in Eq. \ref{KK}), and `aliasing' in the numerical Fourier transform required to relate $S_s(t)$ to $\chi_s''(\omega)$~\cite{B}. 

The results of this paper offer easy tests of the fluctuation-dissipation theorem, that avoids these problems. If Eq. \ref{main} (or Eq. \ref{main2}) is not confirmed by experiment, then the theorem has broken down. In addition to knowledge of the sample shape, this test requires only two `raw' experimental quantities: the noise autocorrelation function $S_s(t)$, which does not have to be on an absolute scale, and the static susceptibility $\chi_s$ -- neither of which have to be corrected for demagnetizing factor.  Similarly if the $1/(1 + N\chi_i)^2$ suppression of the amplitude of $S_s(\omega)$ is not observed, then the theorem has again broken down.  One therefore has easy and robust ways of studying the breakdown of the fluctuation--dissipation theorem as a function of $N$, which can be used to distinguish bulk from surface effects. 

It should be noted that these results strictly apply to systems close to equilibrium such that a full relaxation between true equilibrium states can be observed. This is not necessarily the case for spin glasses in their glassy states~\cite{Ocio, Reim, spinglass}. However, generalisations of the fluctuation-dissipation theorem are relevant to these cases~\cite{efftemp, CugEPL, Cugliandolo} and it would be interesting to understand the present results in that context.  

It is finally worth noting that, using $\langle M^2\rangle_{s/i} =k T \chi_{s/i}/\mu_0 V$ the general result can also be written in the form:  
\begin{equation}
\frac{\tau_s}{\tau_i} = \frac{\chi_s}{\chi_i} 
= \frac{\langle M^2\rangle_s}{\langle M^2\rangle_i},
\end{equation}
a proportionality between mean relaxation time $\tau_s$ and  mean square `displacement', $\langle M^2\rangle_s$.  This is reminiscent of the relationship between time and mean square displacement in Brownian motion, that played a key role in the development of fluctuation theory~\cite{Einstein}.

\acknowledgments{It is a pleasure to thank P. Holdsworth, S. Giblin, C. Paulsen, E. Lhotel and F. Morineau for useful comments and discussion, and S. Giblin for alerting me to Ref. \cite {Finger}. }


\begin{thebibliography}{99}
\bibitem{note} The experimental observable is the magnetic moment ($=$ volume $V$ times magnetization) that is detected through its associated field. In this paper  $S(t)$  is defined as $\mu_0 V$ times the magnetization autocorrelation function where $\mu_0$ is the vacuum permeability.
\bibitem{even} 
For example, both $S(t)$ and $R(t)$ could be bilateral exponential functions, $\sim e^{-|t|}$, in which case $S(\omega)$ will be a Lorentzian function.
\bibitem{st} Fourier transform of Eq. \ref{pink} shows that $$S_i(t)=\frac{3 \chi_i G_{3,9}^{7,3}\left(\frac{t^6}{46656}|
\begin{array}{c}
 \frac{1}{12},\frac{1}{3},\frac{7}{12} \\
 0,\frac{1}{12},\frac{1}{3},\frac{1}{3},\frac{7}{12},\frac{2}{3},\frac{5}{6},\frac{1}{6},\frac{1}{2} \\
\end{array}
\right)}{16 \pi ^{7/2}}$$
where $G$ is the Meijer G-function.
\bibitem{note3} 
An alternative would involve deriving a transformation $S_s(t) \rightarrow S_i(t)$ that does not assume the fluctuation--dissipation theorem, but that seems to be a more challenging problem.
\end{thebibliography}
\end{document}